\documentclass[aps,twocolumn,showpacs]{revtex4}
\usepackage{graphicx}
\begin{document}

\title{Ambipolar Cu- and Fe-Phthalocyanine single-crystal field-effect transistors}
\author{R. W. I. de Boer, A. F. Stassen, M. F. Craciun, C. L. Mulder, A. Molinari, S. Rogge, and A. F. Morpurgo}
\affiliation{Kavli Institute of Nanoscience, Faculty of Applied
Sciences, Delft University of Technology, Lorentzweg 1, 2628 CJ
Delft, the Netherlands}
\date{\today}

\begin{abstract}

We report the observation of ambipolar transport in field-effect
transistors fabricated on single crystals of Copper- and
Iron-Phthalocyanine, using gold as a high work-function metal for
the fabrication of source and drain electrodes. In these devices,
the room-temperature mobility of holes reaches 0.3 cm$^2$/Vs in
both materials. The highest mobility for electrons is observed for
Iron-Phthalocyanines and is approximately one order of magnitude
lower. Our measurements indicate that these values are limited by
extrinsic contact effects due to the transistor fabrication and
suggest that considerably higher values for the electron and hole
mobility can be achieved in these materials.

\end{abstract}


\maketitle

Most of the organic transistors that have been fabricated until
now have shown unipolar conduction, i.e. they function for only
one polarity of the gate voltage \cite{Dimitrakopoulos02}. Only
very recently, ambipolar device operation has been reported in
Field-Effect Transistors (FETs) based on a few different
\textit{single-component} \cite{note1} organic molecular
semiconductors \cite{Meijer03,Chesterfield03,Kunugi04,Yasuda05}.
Some of these ambipolar devices have been used to realize inverter
circuits that demonstrate the possibility of implementing CMOS
technology \cite{Anthopoulos04}. Since CMOS technology is the most
efficient strategy towards the fabrication of integrated circuits
with better noise margins (i.e. insensitivity to the spread in the
parameters of the individual devices) and consuming less power
\cite{Meijer03,Crone00}, this represents a considerable step
forward in the field of plastic electronics. Particularly
important is that the implementation of plastic CMOS integrated
circuits can be realized using a single molecular semiconductor
rather than two different materials, since this has the potential
to considerably simplify the device fabrication processes. So far,
however, the number of single-component molecular semiconductors
for which ambipolar FET operation has been reported is rather
limited.

In this paper we report the investigation of field-effect
transistors fabricated on single-crystals of Copper- and
Iron-Phthalocynanines (Cu- and Fe-Pc's), using gold as a high work
function material for the fabrication of the source and drain
electrodes. Electrical measurements performed in high-vacuum
demonstrate the occurrence of both hole and electron conduction.
From the electrical characteristics of these first organic
single-crystal FET we extract room-temperature mobility values for
holes reaching 0.3 cm$^2$/Vs for both materials. The typical
electron mobility values are different for FePc and CuPc. In FePc
the electron mobility reaches 0.03 cm$^2$/Vs whereas, in CuPc it
is typically lower than 10$^{-3}$ cm$^2$/Vs. These values are
limited by device non-idealities of extrinsic origin, which
originate from difficulties in the FET assembly and which clearly
manifest themselves in the transistor electrical characteristics.
We expect that, upon improving the device assembly procedure,
considerably higher electron and hole mobility can be reached in
Copper- and Iron-Phthalocyanine ambipolar FETs.

The devices used in our studies have been fabricated by means of
electrostatic bonding of very thin ($\sim 1 \ \mu$m thick),
flexible crystals to a substrate onto which the FET circuitry had
been previously fabricated. The details of the technique and of
the crystal growth are similar to those used for the fabrication
of tetracene and rubrene single-crystal FETs \cite{DeBoer04}. From
X-ray diffraction measurements we find that the structure of both
CuPc and FePc crystals, which were grown from vapor phase
\cite{Laudise98}, is the known $\beta$-form \cite{Gould96}. All
crystals used in this work are needle-shaped and their long
direction is aligned along the crystalline $b$-axis.

\begin{figure}[t]
\centering
\includegraphics[width=8.5cm]{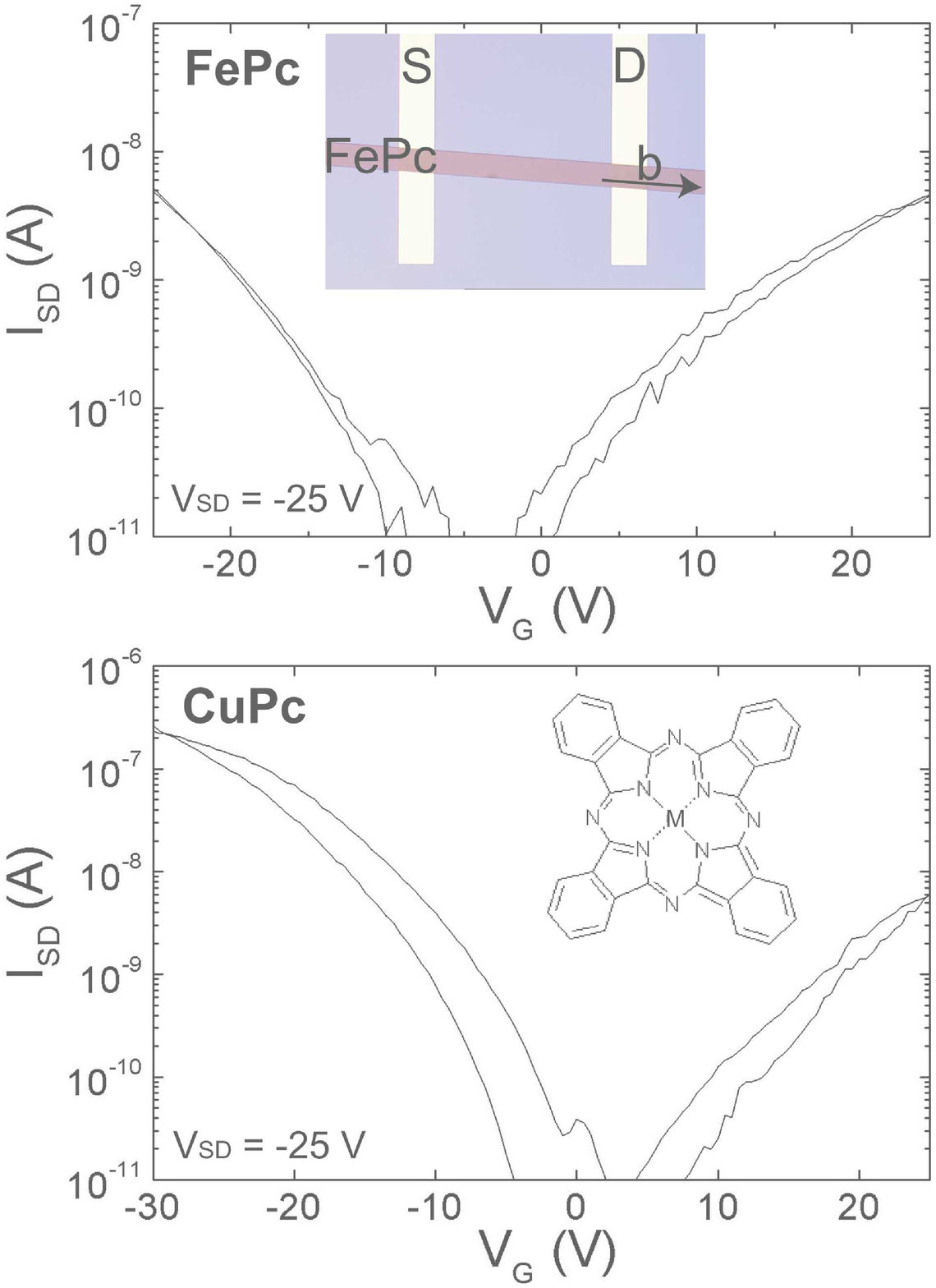}
\caption{Examples of transfer characteristics illustrating
ambipolar transport in FePc and CuPc single-crystal FETs with gold
electrodes. At negative $V_G$ the hole-current is visible, and at
positive $V_G$ there is a clear electron-current. The inset in the
upper plot shows a microscope picture of an FePc FET, with a
distance of $500 \ \mu$m between the source (S) and the drain (D).
The arrow indicates the crystalline $b$-axis. The inset in the
lower plot shows the molecular structure of MPc's.
\label{gatesweep}}
\end{figure}

Contrary to what happens for rubrene or tetracene single crystals
\cite{DeBoer04}, for which the width of very thin crystals can
easily exceed 1 mm, we have found that the width of thin and
flexible CuPc and FePc single crystals that are capable of bonding
to rigid substrates is typically smaller than 100 $\mu$m (see
inset in Fig. \ref{gatesweep}). In addition, very thin crystals
are fragile and easily crack along the crystalline $b$-axis. These
properties of the CuPc and FePc crystals substantially increase
the difficulties involved in the crystal handling and in the
device assembly, often resulting in damage to the crystal or in a
poorer adhesion of the crystals to the substrate and to the
contacts. In spite of these difficulties, we have successfully
fabricated and measured a large number of FETs on FePc and CuPc.
All of the devices fabricated using FePc and approximately half of
the devices fabricated using CuPc exhibited ambipolar transport.

The electrical characterization of the devices has been performed
in a vacuum chamber (pressure $\approx 10^{-7}$ mbar) and in the
dark. All the data shown here were taken in a two terminal
configuration, using a HP4156 parameter analyzer, at room
temperature. We found that heating of the device \textit{in-situ}
up to a temperature of 450 K for a period of several hours results
in a clear improvement of the electrical characteristics (e.g.
reduction of hysteresis) and in a substantial enhancement of the
electron current. This behavior has been observed before and is
attributed to thermally-induced de-sorption of oxygen and water
vapor present in the crystals \cite{Meijer03,DeLeeuw97} and acting
as traps for electrons \cite{Zeis05}. We have also noticed that
heating the device for much longer periods (days) results in
further smaller improvements of the transistor characteristics,
e.g. a factor-of-two enhancement of the mobility of both electrons
and holes. This phenomenon is likely to originate from the
lowering of the contact resistance due to a better adhesion
between crystal and electrodes, or from the annealing of
structural defects.

Fig. \ref{gatesweep} shows logarithmic plots of the source-drain
current $I_{SD}$ measured as a function of gate voltage $V_G$, for
a fixed value of source-drain voltage $V_{SD}$, in a FePc and in a
CuPc device. It is apparent that the measured current increases
with increasing $V_G$ for both negative and positive polarity.
This demonstrates the occurrence of ambipolar transport in these
materials. In general, we observe that at large gate voltage, and
under comparable biasing conditions, the electron current is
normally lower than the hole current (approximately one-to-two
orders of magnitude).

\begin{figure}[t]
\centering
\includegraphics[width=8.5cm]{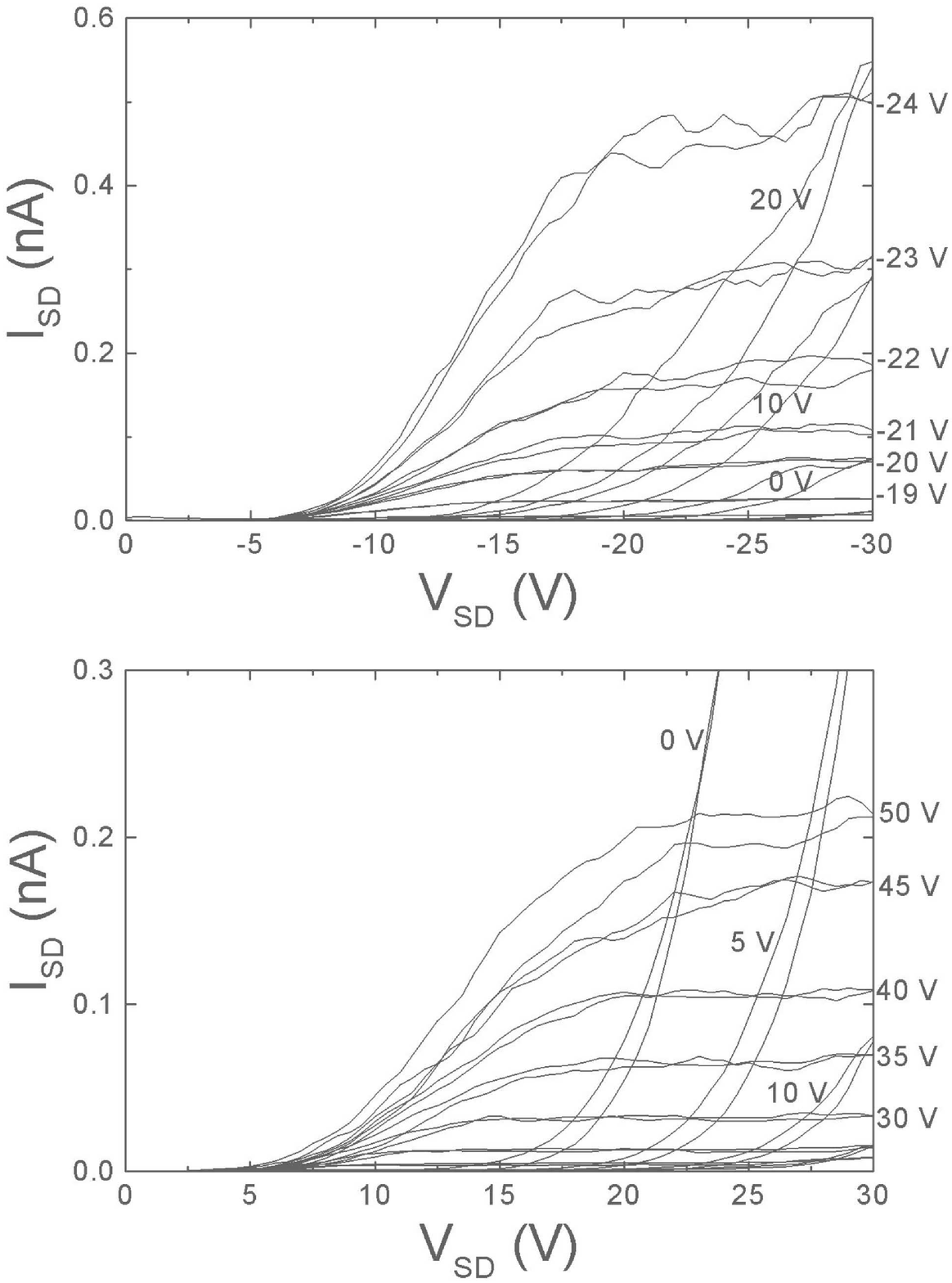}
\caption{Current-voltage characteristics of a FePc single-crystal
FET, demonstrating ambipolar operation. The $V_G$-values are
indicated in the plots. \label{SDsweeps}}
\end{figure}

The transistor characteristics of a FePc device measured by
sweeping the source-drain voltage at different, fixed values of
$V_G$ are shown in Fig.~\ref{SDsweeps}. These measurements
illustrate that the typical transistor behavior is observed for
both negative and positive gate voltage and confirm the occurrence
of ambipolar operation. Overall, the observed behavior is very
similar to that exhibited by recently reported, ambipolar organic
FETs, realized on thin-films of different organic semiconductors
\cite{Meijer03}.

\begin{figure}[t]
\centering
\includegraphics[width=8.5cm]{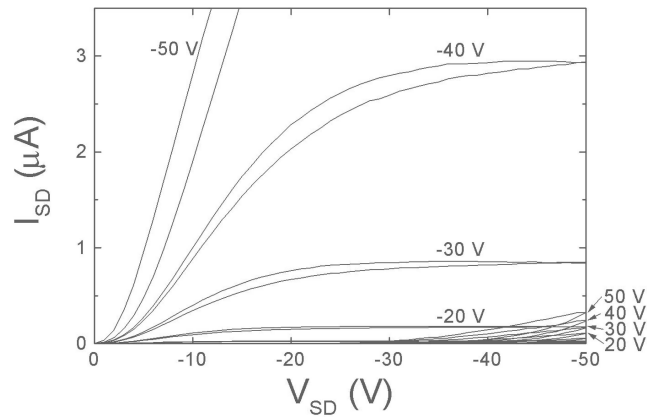}
\caption{Current-voltage characteristics of one of the best FePc
single-crystal FETs, with $\mu_{\mathrm{hole}} = 0.3$ cm$^2$/Vs.
Even in the best devices a clear non-linearity at low $V_{SD}$ is
observed that indicates the presence of contact non-idealities.
\label{SDsweepsbest}}
\end{figure}

The mobility of electrons and holes in different FETs has been
estimated at room temperature from the linear regime of the device
operation. We find a rather large spread in the measured mobility
values. For FePc hole and electron mobilities were found in the
range $10^{-3} - 3 \nolinebreak \cdot \nolinebreak 10^{-1}$
cm$^2$/Vs and $5 \cdot 10^{-4} - 3 \cdot 10^{-2}$ cm$^2$/Vs,
respectively. For CuPc, the measured values for the hole mobility
are in the same range as in FePc, whereas the values for electron
mobility are typically lower ($10^{-5} - 10^{-3}$ cm$^2$/Vs). We
attribute the large spread of mobility to the presence of contact
effects that are visible in the electrical characteristics of even
the best devices (see Fig. \ref{SDsweepsbest}). In this regard,
the electrical behavior of these FePc and CuPc FETs is different
from that observed in tetracene \cite{DeBoer03} and rubrene
\cite{Podzorov03,Podzorov04,Stassen04} single-crystal FET, which
exhibit a considerably smaller spread in mobility values.
Nevertheless, the best mobility values that we have found for both
electron and holes already compare well to the highest mobility
values reported for ambipolar thin-film organic FETs.

Since contact limitations in FePc and CuPc single-crystal FETs are
visible for both holes and electrons even though the work function
of gold is ideally aligned to the HOMO of the molecules, we
conclude that these contact effects are largely of extrinsic
origin, due to the FET assembly process (e.g. imperfect adhesion
of the crystal onto the metal contact). Note, however, that
whereas the values for the hole mobility are approximately the
same for the two different phthalocyanines, the mobility of
electrons is systematically higher in FePc than in CuPc. This is
consistent with the HOMO-LUMO gap of FePc being smaller than that
of CuPc \cite{Liao01}, which allows a more efficient electron in
injection in FePc than in CuPc. Thus, in short, the data suggest
that even though extrinsic contact effects are present and limit
the measured value of the hole mobility to 0.3 cm$^2$/Vs, the
electron current is also significantly limited by the intrinsic
misalignment between the gold work function and the molecular
level for electron transport in the material \cite{Meijer03}.

In comparing the behavior of our single-crystal FETs with
thin-film FETs made with the same materials, it is interesting to
note that ambipolar transport in CuPc thin-film transistors has
been observed only very recently in devices using Calcium
electrodes \cite{Yasuda05}. Even though gold contacted CuPc
thin-film FETs have been extensively studied in the past
\cite{Bao96}, ambipolar operation has never been reported in those
devices. This suggest that the structural quality of the organic
semiconductor (a major difference between thin-film and
single-crystal devices) is also a relevant factor for the
observation of ambipolar FET operation \cite{Zeis05}.

Finally, together with previous experiments on thin-film FETs of
different phthalocyanines in which both p- and n-type operation
has been observed depending on the atmosphere in which the devices
are operated \cite{Guillaud90,Tada00}, our results suggest that it
will be possible to observe ambipolar transport in many different
phthalocyanine-based semiconductors. This is of interest for
fundamental investigations, since the molecular orbitals occupied
by the charge carriers in many different phathalocyanines are
different, whereas the structure of the molecules (as well as that
of their crystals) is essentially identical. For instance, in CuPc
as well as in other metal-Pc's (e.g. MgPc or ZnPc), the
field-induced electrons occupy orbitals that are centered on the
Pc ligands (the doubly-degenerate $2e_g$ orbital). In FePc (and
similarly in CoPc), on the contrary, electrons occupy orbitals
mainly centered on the metal atom \cite{Liao01}. Thus, the
comparative studies of electron transport in different metal-Pc
single-crystal FETs enables the investigation of the relation
between the conduction properties of organic semiconductors and
the electronic properties of the orbitals of the constituent
molecules. For these studies, we are working to improve of the
quality of MPc single-crystal FETs.

We acknowledge H. Kooijman for X-ray diffraction measurements, and
C. Jordanovska for technical assistance. This work was financially
supported by FOM. The work of AFM is part of the NWO
Vernieuwingsimpuls 2000 program.

\end{document}